\documentstyle[aps]{revtex}

\input amssym.def
\input amssym

\begin{document}

\title{ Riemann normal coordinates, Fermi reference system  and the geodesic
deviation equation }

\author{Alexander I. Nesterov
\thanks{Departamento de F\'\i sica, Universidad de Guadalajara,
Guadalajara, Jalisco, M\'exico;
e-mail: nesterov@udgserv.cencar.udg.mx}}



\maketitle


\begin{abstract}
We obtain the integral formulae for computing the tetrads and metric
components in Riemann normal coordinates and Fermi coordinate system of
an observer in arbitrary motion. Our approach admits essential enlarging
the range of validity of these coordinates. The results obtained are
applied to the geodesic deviation in the field of a weak plane
gravitational wave and  the computation of plane-wave metric in Fermi
normal coordinates.
\end{abstract}



\newpage
\section{INTRODUCTION}

The well known Riemann normal coordinates satisfy the conditions
$g_{\mu\nu}|_p =\eta_{\mu\nu}$ and $\Gamma^\mu_{\nu\lambda}|_p=0$ at a point
$p$, the origin of the coordinate system. In these cordinates the metric is
presented in powers of the canonical parameter of the geodesics out from
$p$. In 1922 Fermi \cite{Fermi} showed that it is possible to
generalize the Riemann coordinates in such a way that the Christoffell
symbols vanish along given any curve in a Riemannian manifold, leaving the
metric there rectangular. For a geodesic curve Misner and Manasse \cite{MM}
introduced a Fermi normal coordinates, which satisfy Fermi conditions along
a given geodesic and calculated the first-order expansion of the connection
coefficients and the second-order expansion of the metric in powers of
proper distance normal to the geodesic. Their construction is a special
case of Fermi coordinates defined by Synge \cite{Synge}. The last ones form
a natural coordinate system for a nonrotating accelerated observer.

In 1973 Misner, Thorn and Wheeler (MTW) \cite{MTW} defined  a Fermi
coordinates for an accelerated and rotating observer and calculated the
first-order expansion of the metric. Extending MTW work Mitskievich and
Nesterov \cite{MN3} and later Ni and Zimmerman \cite{NZ} obtained the
second-order expansion of the metric and the first-order expansions of the
connection coefficients. Li and Ni \cite{LN1,LN2} derived the
third-order expansion of the metric and the second-order expansions of the
connection coefficients.

The size of neighbourhood $V$, where Riemann and Fermi normal coordinates
do not involve singularaties, is determined by  conditions
\begin{eqnarray}
u\ll\min \left\{ \frac{1}{| \stackrel{0}R_{\alpha\gamma\delta\beta}
|^{1/2}},~ \frac{| \stackrel{0}R_{\alpha\gamma\delta\beta}
|} {| \stackrel{0}R_{\alpha\gamma\delta\beta,\mu}| }
\right\},  \nonumber
\end{eqnarray}
where $u$ is a canonical parameter along the geodesics from the origin of
coordinates. The first condition
$u \ll |\stackrel{0}R_{\alpha\gamma\delta\beta} |^{-1/2}$ determines the
size of $V$ where the curvature has not yet caused geodesics to cross each
other.  The second condition determines the domain where the curvature does
not change essentially.  For instance, for gravitational waves with
wavelength $\lambda$ the Riemann tensor is $\sim A\exp(i kx)/{\lambda^2}$,
where $A$ is the dimensionless amplitude. This yields \[ u\ll\min \{
\lambda/\sqrt{A},\quad  \lambda\}.  \] Generally it is assumed $A\leq
10^{-18}$. This means that the size of $V$ is restricted by
$u_0\ll\lambda$. So the application of Riemann (or Fermi) coordinates to
the modern experiments may be very restrictive since $\lambda$ is often
supposed being in the order of 300 km. Thus for enlarging the range of
validity by a factor $1/\sqrt A$ (which is about $10^9$ in our example) it
is necesary to take into account derivatives of any order of the Riemann
tensor. Some years ago Marzlin \cite{KPM} has considered Fermi coordinates
for weak gravitational field $g_{\mu\nu}=\eta_{\mu\nu} + h_{\mu\nu}, \;
|h_{\mu\nu}|\ll 1$ and dervived the metric as a Taylor expansion valid to
all orders in the geodesic distance from the world line of Fermi observer.

Here we develop a new approach to Riemann and Fermi coordinates, which
admits enlarging the range of validity of these coordinates up to
$u \ll | \stackrel{0}R_{\alpha\gamma\delta\beta}|^{-1/2}$. The point is to
use the integral formulae. Recently  this approach has been used to
calculate a geometric phase shift for a light beam propagating in the field
of a weak gravitational wave \cite{MN1,MN2}.

The paper is organized as follows. In Section 2 we obtain the integral
formulae for computing the tetrad and metric components in Riemann normal
and Fermi coordinates.  In Section 3 we compute the plane-wave metric in
Fermi normal coordinates and analyse the geodesic deviation equation for
this case.

In this paper we use the space-time signature $(+,-,-,-)$; Greek indices
run from $0$ to $3$, Latin $a,b,c$ from $1$ to $2$ and  $i,j,k$ from $1$ to
$3$.

\section{GENERAL RESULTS}

\subsection{Riemann normal coordinates}

For arbitrary point $p_{0}$ in some neighbourhood $V(p_{0})$ there
is the unique geodesic $\gamma (u)$ connecting $ p_{0}$ and $ p$
which, using  exponential mapping, we may write as $p=\exp_{\gamma(u)}(u\xi)$, where $\xi=\partial/\partial u=\xi^\alpha
e_{(\alpha)}$, $u$ being the canonical parameter while basis is parallely
propagated along $\gamma(u)$.  The Riemann normal coordinates with the
 origin at the point $p_{0}$, are defined as $X^\alpha=u\xi^\alpha$,
where $u\xi=\exp^{-1}_{\gamma(u)} (p)$. The following basic equations
summarize the most important properties of the Riemannian normal
coordinates:
\begin {eqnarray}
\bf e_{(\alpha)}( \em p_{0})=(\partial/\partial{X^\alpha})_{ \em p_{0}}=
({\Lambda_\alpha}^\beta \partial/\partial{x^\beta})_{{ p}_0} ,
~{\Lambda_\alpha}^\beta= (\partial{x^\beta} /\partial{X^\alpha})_{ \em p_0}\\
\stackrel{0}{\;\;\Gamma^\alpha}_{\beta\gamma}=0,~
\Gamma^\alpha_{\beta\gamma}X^\beta X^\gamma=0,~~
\Gamma^\alpha_{\beta\gamma}=-\frac{2}{3}
\stackrel{0}{\;R^\alpha}_{(\beta\gamma)\delta}X^\delta + \cdots,    \\
g_{\alpha\beta}=\eta_{\alpha\beta}+\frac{1}{3}\stackrel{0}
R_{\alpha\gamma\delta\beta}X^\gamma X^\delta+\frac{1}{3!}
\stackrel{0}R_{\alpha\gamma\delta\beta,\mu}X^\gamma X^\delta X^\mu
 \nonumber  \\
 +\frac{1}{5!}\left(6\stackrel{0}R_{\alpha
\gamma\delta\beta,\lambda,\mu}
+\frac{16}{3}\stackrel{0}R_{\alpha\gamma\delta\rho}
\stackrel{0}{R^\rho}_{\lambda\mu\beta}\right)
X^\gamma X^\delta X^\lambda X^\mu  + \cdots
\end{eqnarray}

The size of neighbourhood $V$, where these coordinates do not involve
singularaties, is determined by  conditions
\begin{equation}
u\ll\min \left\{ \frac{1}{| \stackrel{0}R_{\alpha\gamma\delta\beta}
|^{1/2}},~ \frac{| \stackrel{0}R_{\alpha\gamma\delta\beta}
|} {| \stackrel{0}R_{\alpha\gamma\delta\beta,\mu}| }
\right\}.
\end{equation}
The first condition $u \ll | \stackrel{0}R_{\alpha\gamma\delta\beta}
|^{-1/2}$ determines the size of $V$ where the curvature has not yet
caused geodesics to cross each other. The second condition determines the
domain where the curvature does not change essentially.

For studing Riemann and Fermi normal coordinates one useful technique is
the geodesic deviation equation \cite{MM}. Further we apply it for
our computation. Let us consider a point $p_{0}$ and a frame $\{{\bf
e}_{(\alpha)}(p_{0})\}$ in this point.  For arbitrary geodesic $\gamma(u)$
out from $p_{0}$ the equation of geodesic deviation has the following form
\begin{equation}
\nabla_{\xi}^2\eta+{\sf R}(\xi,\eta)\xi=0,
\label{dev}
\end{equation}
where $\xi=\partial/\partial u$ is tangent vector to $\gamma(u)$, $\eta$ is
vector of deviation and $\sf R$ is operator of curvature. We suppose that
the basis $\bf e_{(\alpha)}$ is parallely propagated along
$\gamma(u)$.  In tetrad components the equation (\ref{dev}) takes the form
\begin{equation}
\frac{d^2 \eta^{(\alpha)}}{du^2}={R^{(\alpha)}}_{(\beta)(\gamma)(\delta)}
\xi^{(\beta)}\xi^{(\gamma)}\eta^{(\delta)}
\label{dev1}
\end{equation}
where $\eta^{(\alpha)}= \eta^{\beta}e_\beta^{(\alpha)}$, etc.

In Riemann normal coordinates there are natural deviation vectors
 $\eta_{\alpha}=u\delta^\beta_\alpha\partial/\partial X^\beta$ satisfying
Eq.(\ref{dev}) \cite{MM}.  Then (\ref{dev}) yields \begin{equation}
\frac{d^2 e^{(\alpha)}_\beta}{d u^2 }
+\frac{2}{u}\frac{d e^{(\alpha)}_\beta}{d u }=
e^{(\alpha)}_\lambda {R^\lambda}_{\gamma\delta\beta}\xi^\gamma
\xi^\delta.
\label{Int}
\end{equation}

We solve this equation by succesive  approximations \cite{Arf}.
Let us write (\ref{Int}) in the equivalent integral form
\begin{eqnarray}
e^{(\alpha)}_{\beta}= \delta^\alpha_\beta +  \frac{1}{u}{\int_0^u
d\tau\int_0^\tau d\tau '\tau '\!{e^{(\alpha)}_\lambda}
{R^\lambda}_{\gamma\delta\beta}\xi^\gamma \xi^\delta},
\label{It0}
\end{eqnarray}
where we assume $e^{(\alpha)}_{\beta}= \delta^\alpha_\beta$ at the origin
of Riemann coordinates. Now we start with
\[
e^{(\alpha)}_{\beta}\approx   e^{(\alpha)}_{0\beta} =
\delta^\alpha_\beta.
\]
To improve this first approximation we feed back
$e^{(\alpha)}_{0\beta}$ into the integral (\ref{It0}) getting
\begin{eqnarray}
e^{(\alpha)}_{1\beta}
=\delta^\alpha_\beta +\frac{1}{u}{\int_0^u
d\tau\int_0^\tau d\tau '\tau '\!
{R^\alpha}_{\gamma\delta\beta}\xi^\gamma \xi^\delta},
\end{eqnarray}
Repeteang this process by substituting the new $e^{(\alpha)}_{i\beta}$
back into Eq. (\ref{It0}), we find our solution $e^{(\alpha)}_\beta$ as
infinite series
\begin{eqnarray}
&& e^{(\alpha)}_\beta= \sum_{n=0}^{\infty}\!\!
\stackrel{~~(n)(\alpha)}{e_\beta}, \label{ser} \\
&& \stackrel{~~(0)(\alpha)}{e_\beta}= \delta^\alpha_\beta \nonumber \\
&& \stackrel{~~(1)(\alpha)}{e_\beta}=\frac{1}{u}{\int_0^u
d\tau\int_0^\tau d\tau '\tau '\!\stackrel{~~(0)(\alpha)}{e_\lambda}
{R^\lambda}_{\gamma\delta\beta}\xi^\gamma \xi^\delta},\nonumber \\
&& \stackrel{~~(2)(\alpha)}{e_\beta}=\frac{1}{u}{\int_0^u
d\tau\int_0^\tau d\tau '\tau '\!\stackrel{~~(1)(\alpha)}{e_\lambda}
{R^\lambda}_{\gamma\delta\beta}\xi^\gamma \xi^\delta},\nonumber \\
&& \stackrel{~~(n+1)(\alpha)}{e_\beta}=\frac{1}{u}{\int_0^u
d\tau\int_0^\tau d\tau '\tau '\!\stackrel{~~(n)(\alpha)}{e_\lambda}
{R^\lambda}_{\gamma\delta\beta}\xi^\gamma \xi^\delta}.\nonumber
\end{eqnarray}
Noting that the order of magnitude of functions
$\stackrel{~~~(n)(\alpha)}{e_\beta}$ is given in the neighbourhood $V$ by
\[
\stackrel{~~~(n)(\alpha)}{e_\beta}\leq
\frac{{|R_{\alpha\gamma\delta\beta}X^\gamma X^\delta
|}^n_{\max}}{(2n+1)!!},
\]
we conclude that the radius of convergence of series (\ref{ser}) is
determined by
\[
u_0\simeq \left\{ \frac{1}{\left|
{R_{\alpha\gamma\delta\beta}}\right|_{\max}^{1/2}}\right\}.
\]

Now using (\ref{Int}-\ref{ser}) one obtains for the components of
orthonormal tetrad in Riemann coordinates the following integral formula:
\begin{equation}
e^{(\alpha)}_\beta=\delta^\alpha_\beta+
\frac{1}{u}{\int_0^u d\tau\int_0^\tau d\tau '\tau
'{R^\alpha}_{\gamma\delta\beta}\xi^\gamma \xi^\delta}+{\cal O}(R^2).
\label{int}
\end{equation}

{\it Remark 2.1.} Integrating by parts one can write (\ref{int}) as
\begin{equation}
e^{(\alpha)}_\beta=\delta^\alpha_\beta+
{\int_0^u d\tau\int_0^\tau d\tau '
{R^\alpha}_{\gamma\delta\beta}\xi^\gamma \xi^\delta}
-\frac{2}{u}{\int_0^u d\tau\int_0^\tau d\tau '\int_0^{\tau '} d\tau ''
{R^\alpha}_{\gamma\delta\beta}\xi^\gamma \xi^\delta}+ {\cal O}(R^2).
\end{equation}
This form is more convenient for the comparison with the results
obtained in Fermi coordinates (see Subsection B). For the covariant
derivative of the tetrad  we obtain the following useful formula (see
Appendix A)
\begin{equation}
\nabla_{\partial_\lambda}{e_{(\nu)}}^\mu=
-\frac{1}{u}\int_0^u {R^\mu}_{\nu \lambda \rho}{\xi^\rho}\tau d\tau
+ {\cal O}(R^2).
\label{exp1}
\end{equation}

With the aid of (\ref{int}), it is straightforward to derive the
metric in Riemann coordinates by using
$g_{\alpha\beta}=\eta_{\mu\nu}e^{(\mu)}_\alpha e^{(\nu)}_\beta$. The result
is
\begin{equation}
g_{\alpha\beta}=\eta_{\alpha\beta} +
\frac{2}{u}{\int_0^u d\tau\int_0^\tau d\tau '\tau '
R_{\alpha\gamma\delta\beta}\xi^\gamma \xi^\delta}
+{\cal O}(R^2),
\label{eq13}
\end{equation}
and can be rewritten by using Taylor expansion as
\[
g_{\alpha\beta}=\eta_{\alpha\beta}+\frac{1}{3}\stackrel{0}
R_{\alpha\gamma\delta\beta}X^\gamma X^\delta+\frac{1}{3!}
\stackrel{0}R_{\alpha\gamma\delta\beta,\mu}X^\gamma X^\delta X^\mu
+\frac{6}{5!} \stackrel{0}R_{\alpha \gamma\delta\beta,\lambda,\mu}
X^\gamma X^\delta X^\lambda X^\mu  + \cdots  ,
\]
that agrees with Eq.(3).

	In fact, we are interested in expressions which should involve (in
the right-hand sides) the arbitrary chosen initial non-Riemann coordinates
only. Then
\begin{equation}
e^{(\alpha)}_\beta=\delta^ \alpha_\beta+\frac{1}{u}{\int_0^u
d\tau\int_0^\tau d\tau '\tau
'{R^\lambda}_{\gamma\delta\sigma}\frac{\partial X^\alpha}{\partial
x^\lambda} \frac{\partial x^\sigma}{\partial X^\beta}\frac{\partial
x^\gamma}{\partial X^\kappa}\frac{\partial x^\delta}{\partial
X^\rho}\xi^\kappa \xi^\rho}+{\cal O}(R^2).
\label{eq14}
\end{equation}
The transformation from arbitrary coordinate system to Riemann one takes
the form $x^\mu=x^\mu(p_{0})+\Lambda^\mu_\nu X^\nu +{\cal O}(\Gamma)$ (or
$x^\mu=x^\mu(p_{0})+\Lambda^\mu{}_\nu\xi^\nu u  +{\cal O}(\Gamma)$ ) and the
equations (\ref{eq13}), (\ref{eq14}) can be written  as
\begin{eqnarray}
e^{(\alpha)}_\beta=\delta^ \alpha_\beta+\frac{1}{u}{\int_0^u
d\tau\int_0^\tau d\tau '\tau
'{R^\lambda}_{\gamma\delta\sigma}\Lambda^{-1\alpha}_\lambda
{\Lambda_\kappa} ^\gamma {\Lambda_\rho} ^\delta{\Lambda_\beta}
^\sigma\xi^\kappa \xi^\rho}+ \mbox{\scriptsize$\cal O$}(R), \\
g_{\alpha\beta}=\eta_{\alpha_\beta}+\frac{2}{u}{\int_0^u d\tau\int_0^\tau
d\tau '\tau '{R_\lambda}_{\gamma\delta\sigma}\Lambda^\lambda _{\alpha}
{\Lambda_\kappa} ^\gamma {\Lambda_\rho} ^\delta{\Lambda_\beta}
^\sigma\xi^\kappa \xi^\rho}+ \mbox{\scriptsize$\cal O$}(R),
\end{eqnarray}
where
$R^\lambda{}_{\gamma\delta\beta}=R^\lambda{}_{\gamma\delta\sigma}
\Big (x^\mu(p_{0}) + \Lambda^\mu_\nu \xi^\nu \tau'\Big )$ while the
integration is performed.

\subsection{Fermi coordinates}

Let us consider the proper system of reference $\bf e_{(\alpha)}$ of a
single observer which moves along a worldline $\gamma$ with an acceleration
and rotation \cite{MTW}

\begin{equation}
\frac{D{\bf e_{(\alpha)}}}{ds}= \bf\Omega \cdot \bf e_{(\alpha)},
\end{equation}
where
\[
\Omega^{\mu \nu}=G^\mu \tau^\nu-G^\nu \tau^\mu+
E^{\mu\nu\gamma\delta}\tau_\gamma \omega_\delta
\]
where $E^{\mu\nu\gamma\delta}$ is the axial tensor of Levi-Civita and
$\tau^\mu$ is the unit tangent vector to the line $\gamma$; $G^\mu$ being
the four-acceleration of observer, $\omega^\delta$ the four-rotation of
the tetrad. Fermi coordinates  are defined as
$X^\mu=\delta^\mu_0s+\delta^\mu_i {\xi^i}u$ (see {\it e.g.} \cite{MM,MTW}).
Here $s$ is proper time along the observer's world line $\gamma$,
$\mbox\boldmath{\xi}$ is a unit vector on $\gamma$ tangent to the spacelike
geodesic starting from $\gamma$, ortogonal to it and parametrized by proper
length $u$. In fact, $\mbox\boldmath{\xi}$ describes the direction in which
such a space-like geodesic goes, and it possesses only spatial components
different from zero (the temporal coordinate is directed along $\gamma$).
In Fermi coordinates, the covariant components of the corresponding
orthonormal tetrad ${e^{(\nu)}_\mu}$ parallel transported along the
spacelike geodesic, are represented as expansions (see Appendix B)
\begin{eqnarray}
{e^{(\mu)}_0}=\delta^\mu_0+\Omega^\mu{}_i
X^i+\frac{1}{2}\stackrel{0~}{R^\mu}_{ij0}X^iX^j
+ \frac{1}{6}\stackrel{0}{R}_{0ij0}\Omega^\mu{}_k
X^i X^jX^k  + \frac{1}{6}\stackrel{0~}{R^\mu}_{ij0,k}X^i X^jX^k+\cdots,
\nonumber   \\
{e^{(\mu)}}_p=\delta^\mu_p+\frac{1}{6}\stackrel{0~}{R^\mu}_{ijp}X^iX^j
+\frac{1}{12}\stackrel{0~}{R^0}_{ijp}\Omega^\mu{}_k X^iX^jX^k
+\frac{1}{12}\stackrel{0~}{R^\mu}_{ijp,k} X^iX^jX^k+\cdots,
\label{exp}
\end{eqnarray}
where quantities of the type of $\stackrel{0}{Q}$ are taken on $\gamma$.
The expansion up to 3d order of the metric and up to 1st order of
the connection coefficients (Christoffel symbols) is given by
\cite{MN3,NZ,LN1,LN2}
\begin{eqnarray}
g_{00}= 1+2\Omega_{0p}X^p + \Omega_{\alpha
k}\Omega^\alpha{}_{p} X^k X^p +\stackrel{0}R_{0kp0}X^kX^p
+\stackrel{0}{R}_{\alpha ij0}\Omega^\alpha{}_kX^i X^jX^k  \nonumber \\
+\frac{1}{3}\stackrel{0}{R}_{0ij0}\Omega_{0k}X^i X^jX^k
+\frac{1}{3}\stackrel{0}{R}_{0ij0,k}X^i X^jX^k + \cdots,\nonumber \\
g_{0i}=\Omega_{ip}X^p +\frac{2}{3}\stackrel{0}R_{ikl0}X^kX^l
+\frac{1}{4}\stackrel{0}R_{ikl0}\Omega_{0m}X^kX^lX^m
+\frac{1}{6}\stackrel{0}R_{0kl0}\Omega_{im}X^kX^lX^m   \nonumber \\
+\frac{1}{6}\stackrel{0}R_{iklm}\Omega^m{}_{n}X^kX^lX^n
+\frac{1}{4}\stackrel{0}R_{ikl0,m}X^kX^lX^m + \cdots, \nonumber \\
g_{ij}=\eta_{ij}+\frac{1}{3}\stackrel{0}R_{ikpj}X^kX^p
+\frac{1}{12}(\stackrel{0}R_{ikl0}\Omega_{jm}X^kX^lX^m
+\stackrel{0}R_{jkl0}\Omega_{im}X^kX^lX^m)  \nonumber \\
+\frac{1}{6}\stackrel{0}R_{iklj,m}X^kX^lX^m + \cdots, \nonumber \\
\Gamma^\mu_{0\nu}={\Omega^\mu}_\nu+(\delta^0_\nu
{\dot{\Omega}^\mu}\thinspace_i+
\delta^0_\nu{\Omega^\mu}_\lambda{\Omega^\lambda}_i
-{\Omega^0}_\nu{\Omega^\mu}_i+\stackrel{0~}{R^\mu}_{\nu i0})X^i
+\cdots, \nonumber   \\
\Gamma^\mu_{ij}= -\frac{2}{3}\stackrel{0~}{R^\mu}_{(ij)k}X^k +\cdots.
\label{eq18}
\end{eqnarray}
For applications it is convenient to write the expansion for the metric
tensor in the following form
\begin{eqnarray}
g_{00}= (1+\mbox{\boldmath{$G X$}})^2
- (\mbox{\boldmath{$\omega\times X$}})^2+\stackrel{0}R_{0kp0}X^kX^p
+\stackrel{0~}{R^l}_{ij0}\epsilon_{lkp}\omega^p X^i X^jX^k  \nonumber \\
+\frac{4}{3}\stackrel{0}{R}_{0ij0}G_k X^i X^jX^k
+\frac{1}{3}\stackrel{0}{R}_{0ij0,k}X^i X^jX^k + \cdots,\nonumber \\
g_{0i}=-(\mbox{\boldmath{$\omega\times X$}})_i
+\frac{2}{3}\stackrel{0}R_{0kpi}X^kX^p
+\frac{1}{4}\stackrel{0}R_{ikl0}G_mX^kX^lX^m
+\frac{1}{6}\stackrel{0}R_{0kl0}\epsilon_{imn}\omega^n X^kX^lX^m
\nonumber \\
+\frac{1}{6}\stackrel{0~}{R^m}_{kli}\epsilon_{mnp}\omega^p X^kX^lX^n
+\frac{1}{4}\stackrel{0}R_{ikl0,m}X^kX^lX^m + \cdots, \nonumber \\
g_{ij}=\eta_{ij}+\frac{1}{3}\stackrel{0}R_{ikpj}X^kX^p
+\frac{1}{12}\big (\stackrel{0}R_{ikl0}\epsilon_{jmn}\omega^n X^kX^lX^m
+\stackrel{0}R_{jkl0}\epsilon_{imn}\omega^n X^kX^lX^m \big )
\nonumber \\
+\frac{1}{6}\stackrel{0}R_{iklj,m}X^kX^lX^m + \cdots,
\nonumber
\end{eqnarray}
where the expression $(\mbox{\boldmath{$\omega\times X$}})_i $ denotes
$\epsilon_{ijk}\omega^j X^k$  and $\mbox{\boldmath{$G X$}}:=-G_i X^j$.

The radius of convergence of series (\ref{eq18}) is determined by
(see Introduction)
\begin{eqnarray}
u\ll u_0 = \min \left\{\frac{1}{|\mbox{\boldmath$G$}|},\quad
\frac{1}{|\mbox{\boldmath$\omega$}|},\quad \frac{1}{|
\stackrel{0}R_{\alpha\gamma\delta\beta}|^{1/2}},\quad \frac{|
\stackrel{0}R_{\alpha\gamma\delta\beta} |} {|
\stackrel{0}R_{\alpha\gamma\delta\beta,i}|} \right\}.\nonumber
\end{eqnarray}

Let us compute the integral formula similar to (\ref{int}). In Fermi
coordinates the deviation vectors are given by \cite{MM}
\[
\eta_{0}=\delta^\mu_0\frac{\partial}{\partial X^\mu},
\quad \eta_{i} = u\delta^\mu_i\frac{\partial}{\partial X^\mu}.
\]
The straightforward calculation leads to
\begin{eqnarray}
e^{(\alpha)}_p=\delta^ \alpha_p+\frac{1}{u}{\int_0^u
d\tau\int_0^\tau d\tau '\tau 'e^{(\alpha)}_\lambda
{R^\lambda}_{ijp}\xi^i\xi^j},\quad  \\
e^{(\alpha)}_0=\delta^ \alpha_0 + \frac{d\stackrel{0\;(\alpha)}{e_0\quad}}
{d u}u+{\int_0^u d\tau\int_0^\tau d\tau '
e^{(\alpha)}_\lambda {R^\lambda}_{ij 0}\xi^i \xi^j}.
\end{eqnarray}
where
\[
\stackrel{0\;(\alpha)}{e_\mu\quad}=\delta^ \alpha_\mu, \quad
\frac{d\stackrel{0\;(\alpha)}{e_\mu\quad}}{d u}
= \delta^0_\mu \Omega^\alpha{}_i\xi^i.
\]
The iterative scheme (\ref{ser}) yields
\begin{eqnarray}
e^{(\alpha)}_\beta=
\sum_{n=0}^{\infty}\!\! \stackrel{~~(n)(\alpha)}{e_\beta}, \quad
\stackrel{~~(0)(\alpha)}{e_\beta}=\delta^ \alpha_\beta
+\delta^ 0_\beta \Omega^\alpha{}_i\xi^i u,
\label{ser01} \\
\stackrel{~~(n+1)(\alpha)}{e_\beta}=\frac{1}{u}{\int_0^u
d\tau\int_0^\tau d\tau '\tau '\!\stackrel{~~(n)(\alpha)}{e_\lambda}
{R^\lambda}_{\gamma\delta\beta}\xi^\gamma \xi^\delta},
\label{It01}
\end{eqnarray}
and the radius of convergence of series (\ref{ser01}) is determined by
conditions
\begin{eqnarray}
u\ll u_0 = \min \left\{\frac{1}{|\mbox{\boldmath$G$}|},\quad
\frac{1}{|\mbox{\boldmath$\omega$}|},\quad
\frac{1}{|\stackrel{0}R_{\alpha\gamma\delta\beta}|^{1/2}} \right\}.\nonumber
\end{eqnarray}

In the first approximation one obtains
\begin{eqnarray}
e^{(\alpha)}_p=\delta^ \alpha_p+\frac{1}{u}{\int_0^u
d\tau\int_0^\tau d\tau '\tau '(\delta^\alpha_\lambda+ \delta^0_\lambda
\Omega^\alpha{}_i\xi^i\tau') {R^\lambda}_{jkp}\xi^j\xi^k} +{\cal O}(R^2),
\quad \label{tetr1} \\
e^{(\alpha)}_0=\delta^ \alpha_0
+ \Omega^\alpha{}_i\xi^i u
+{\int_0^u d\tau\int_0^\tau d\tau '(\delta^\alpha_\lambda+
\delta^0_\lambda \Omega^\alpha{}_i\xi^i\tau')
{R^\lambda}_{jk 0}\xi^j \xi^k} +{\cal O}(R^2).
\label{tetr2}
\end{eqnarray}
Integrating by parts we combine (\ref{tetr1}), (\ref{tetr2})  into one
expression
\begin{eqnarray}
e^{(\alpha)}_\beta=\delta^ \alpha_\beta
+\delta^0_\beta \Omega^\alpha{}_i\xi^i u
+{\int_0^u d\tau\int_0^\tau d\tau '(\delta^\alpha_\lambda+
\delta^0_\lambda \Omega^\alpha{}_i\xi^i\tau')
{R^\lambda}_{jk \beta}\xi^j \xi^k} \nonumber \\
-\frac{2}{u}{\int_0^u d\tau\int_0^\tau d\tau ' \int_0^{\tau'}d\tau''
\delta^p_\beta (\delta^\alpha_\lambda+ \delta^0_\lambda
\Omega^\alpha{}_i\xi^i\tau'') {R^\lambda}_{jkp}\xi^j\xi^k} +{\cal O}(R^2),
\label{eq25}
\end{eqnarray}
and the componets of metric tensor are given by
\begin{eqnarray}
g_{\alpha_\beta}=\eta_{ \alpha_\beta}
+\delta^0_\beta \Omega_{\alpha i}\xi^i u
+\delta^0_\alpha \Omega_{\beta i}\xi^i u
+\delta^0_\alpha \delta^0_\beta\Omega_{\gamma i}
 \Omega^\gamma_j\xi^i \xi^j u^2       \nonumber \\
+(\delta^\mu_\alpha + \delta^0_\alpha \Omega^\mu{}_i\xi^i u)
\left [{\int_0^u d\tau\int_0^\tau d\tau '(\eta_{\mu\lambda}+
\delta^0_\lambda \Omega_{\mu i}\xi^i\tau')
{R^\lambda}_{jk \beta}\xi^j \xi^k} \right. \nonumber \\
- \left.\frac{2}{u}{\int_0^u d\tau\int_0^\tau d\tau ' \int_0^{\tau'}d\tau''
\delta^p_\mu (\eta_{\beta\lambda}+ \delta^0_\lambda \Omega_{\beta
i}\xi^i\tau'') {R^\lambda}_{jkp}\xi^j\xi^k} \right] \nonumber \\
+(\delta^\mu_\beta + \delta^0_\beta \Omega^\mu{}_i\xi^i u)
\left [{\int_0^u d\tau\int_0^\tau d\tau '(\eta_{\mu\lambda}
+ \delta^0_\lambda \Omega_{\mu i}\xi^i\tau') {R^\lambda}_{jk \alpha}\xi^j
\xi^k} \right. \nonumber \\
 -\left. \frac{2}{u}{\int_0^u d\tau\int_0^\tau d\tau '
\int_0^{\tau'}d\tau'' \delta^p_\mu (\eta_{\alpha\lambda}+
\delta^0_\lambda \Omega_{\alpha i}\xi^i\tau'') {R^\lambda}_{jkp}\xi^j\xi^k}
\right ] +{\cal O}(R^2),
\label{g}
\end{eqnarray}
This integral formula agrees with the series (\ref{eq18}).

The transformation from arbitrary coordinate system $\{x^\mu\}$ to Fermi
one  takes the form
\[
x^\mu=x^\mu(X^0)+\Lambda^\mu_i(X^0)X^i + \cdots ,
\]
and in the coordinates  $\{x^\mu\}$ one can write Eq.(\ref{eq25}) as
\begin{eqnarray}
e^{(\alpha)}_\beta=\delta^ \alpha_\beta
+\delta^0_\beta \Omega^\gamma{}_i \Lambda^{-1\alpha}_\gamma\xi^i u
+{\int_0^u d\tau\int_0^\tau d\tau '(\Lambda^{-1\alpha}_\gamma+
\Lambda_\lambda{}^0\Lambda^{-1\alpha}_\gamma \Omega^\gamma{}_i \xi^i\tau')
{R^\lambda}_{\delta\kappa \beta}\Lambda_j{}^\delta\Lambda_k{}^\kappa
\xi^j\xi^k} \nonumber \\
-\frac{2}{u}{\int_0^u d\tau\int_0^\tau d\tau ' \int_0^{\tau'}d\tau''
\delta^p_\beta (\Lambda^{-1\alpha}_\gamma+
\Lambda_\lambda{}^0\Lambda^{-1\alpha}_\gamma \Omega^\gamma{}_i \xi^i\tau'')
{R^\lambda}_{\delta\kappa p}\Lambda_j{}^\delta\Lambda_k{}^\kappa\xi^j\xi^k}
+\mbox{\scriptsize$\cal O$}(R),\nonumber
\end{eqnarray}
where it is used the same notation $R^\alpha{}_{\beta\gamma\delta}$ for the
curvature tensor in the coordinate system $x^\mu$ and
\begin{eqnarray}
R^\alpha{}_{\mu\nu\lambda} = R^\alpha{}_{\mu\nu\lambda}\left(x^\beta(X^0) +
\Lambda^\beta_i(X^0)\xi^i\tau''\right) \nonumber
\end{eqnarray}
while the integration is performed.

\section{Weak plane-wave metric in Fermi normal coordinates}

We compute the plane-wave metric in Fermi normal coordinates for a geodesic
observer in the  weak gravitational-wave field, whose metric tensor is
usually written in synchronous coordinates,
\[
{ds}^2=\eta_{\mu\nu}dx^{\mu} dx^{\nu}+{h}_{ab}dx^a dx^b,
\]
where $a$ and $b$ run from 1 to 2 while
\[
h_{a b}= h_{a b} (t-z),~  h_{2 2}= -h_{1 1} = h_{+},~ h_{1 2}=
h_{\times}.
\]
Using the definition of the Riemann tensor
\[
R_{\mu\nu\lambda\sigma}=\frac{1}{2}(h_{\nu\lambda,\mu,\sigma}+
h_{\mu\sigma,\nu\,lambda}-h_{\mu\lambda,\nu,\sigma}
-h_{\nu\sigma,\mu,\lambda}),
\]
one finds that in the linearized theory non-zero components of Riemann
tensor are
\begin{eqnarray}
R_{3ab3}=R_{0ab0}=-R_{3ab0}=\frac{1}{2}\ddot
h_{ab},\nonumber \\
R_{\mu22\nu}=-R_{\mu11\nu}, \; R_{\mu12\nu}=R_{\mu21\nu} \nonumber,
\end{eqnarray}
dot being derivative with respect to $t$.

The next step is to choose an orthonormal frame along the world line
$\gamma$ of observer. Taking into account the transformation to the Fermi
normal coordinates $x^\mu=X^\mu+{\cal O}(h_{ab})$ and that the timelike
base vector must be the tangent $\partial /{\partial t}$, one obtains
\begin{eqnarray}
{\bf e}_{(0)}|_\gamma ={\partial}/{\partial T}=
{\partial}/{\partial t}, \quad {\bf e}_{(1)}|_\gamma =
{\partial}/{\partial X} = {\partial}/{\partial x} , \nonumber\\
{\bf e}_{(2)}|_\gamma ={\partial}/{\partial Y} = {\partial}/{\partial y},
\quad {\bf e}_{(3)}|_\gamma  = {\partial}/{\partial Z} =
{\partial}/{\partial z} , \nonumber
\end{eqnarray}
where $X^0=T,\; X^1=X,\;X^2=Y,\; X^3=Z$ are setted.

Using (\ref{g}) we compute the metric tensor in the Fermi normal
coordinates. The result is
\begin{eqnarray}
g_{00}=1+H_{ab}X^a X^b, \\
g_{ab}=\eta_{ab}+F_{ab}Z^2, \\
g_{0a}=\frac{1}{2}\left(H_{ab}  +F_{ab}\right)X^b Z, \\
g_{03}=-\frac{1}{2}\left(H_{ab} +F_{ab}\right) X^a X^b , \\
g_{33}=-1+F_{ab} X^a X^b ,   \\
g_{a3}=-F_{ab} X^b Z,
\end{eqnarray}
where
\begin{eqnarray}
H_{ab}=\frac{1}{Z^2}\left(h_{ab}(T-Z)-h_{ab}(T) +
Z {\dot h}_{ab}(T) \right) ,\nonumber\\
F_{ab}=H_{ab} -\frac{2}{Z^3}\int_0^Z
H_{ab}(T,Y){Y}^2 d Y. \nonumber
\end{eqnarray}
It is easy to see that ${H'}_{ab} = - {\dot F}_{ab}$,
$H_{ab}(T,0)=3F_{ab}(T,0)=(1/2){\ddot h}_{ab}(T)$ and
\[
F_{ab} = \frac{1}{2}((H_{ab} - F_{ab})Z)',
\]
here and below dot being derivative with respect to $T$ and prime
derivative with respect to $Z$. The nonvanishing components of
Christoffel symbols are found to be
\begin{eqnarray}
\Gamma^a_{0b}=\frac{1}{2}H'_{ab}Z^2,\quad
\Gamma^a_{00}=H_{ab}X^b -\frac{1}{2}(\dot H_{ab}-H'_{ab})X^b Z,\\
\Gamma^a_{33}=2F_{ab}X^b + F'_{ab}X^b Z,\quad
\Gamma^a_{3b}= -F_{ab}Z - \frac{1}{2} F'_{ab}Z^2 \\
\Gamma^a_{03}=-H_{ab} X^b -H'_{ab}Z X^b, \quad
\Gamma^0_{0a}=\Gamma^3_{0a}= H_{ab}X^b, \\
\Gamma^0_{ab}=\Gamma^3_{ab}= 2F_{ab}Z + \frac{1}{2}F'_{ab}Z^2 ,\quad
\Gamma^3_{a3}=- F_{ab}X^b, \\
\Gamma^0_{00}=\Gamma^3_{00}=\frac{1}{2}\dot H_{ab}X^a X^b,\quad
\Gamma^0_{33}=-\frac{1}{2}F'_{ab}X^a X^b,\\
\Gamma^0_{03} = \Gamma^3_{03}=\frac{1}{2}H'_{ab}X^a X^b,\quad
\Gamma^3_{33}=-\frac{1}{2}F'_{ab}X^a X^b,
\end{eqnarray}
and the computation of the non-zero components of Riemann tensors in
Fermi normal coordinates yields
\begin{eqnarray}
R_{3ab3}=R_{0ab0}=-R_{3ab0} \nonumber \\
=\frac{1}{2}\{\ddot F_{ab}Z^2+2H_{ab}
-(\dot H_{ab}+\dot F_{ab})Z\}.
\end{eqnarray}
This leads to
\begin{eqnarray}
R_{3ab3}=R_{0ab0}=-R_{3ab0}=\frac{1}{2}\ddot h_{ab}(T-Z).
\end{eqnarray}
Thus in Fermi coordinates the curvature is the same function
$(1/2){\ddot h}_{ab}$, but depending on $(T-Z)$.

{\em Plane monochromaric gravitational wave.}
The only nonzero components $h_{ab} = A_{ab}\sin[k(t-z)]$ of wave
amplitudes are
\begin{eqnarray}
h_{11}=-h_{22}= A_{+}\sin[k(t-z)], \nonumber \\
h_{12}=h_{21}= A_{\times}\sin[k(t-z)]. \nonumber
\end{eqnarray}
Computation yields
\begin{eqnarray}
H_{ab}=k^2 A_{ab}\left[ \cos(kT)\frac{kZ-\sin(kZ)}{(kZ)^2}
+ \sin(kT)\frac{\cos(kZ)-1}{(kZ)^2} \right] ,\\
F_{ab}=\left\{ \cos(kT)\left[\frac{kZ
-\sin(kZ)}{(kZ)^2} - 2\frac{\cos(kZ)-1+(kZ)^2/2}{(kZ)^3} \right]   \right.
\nonumber \\ \left.+ \sin(kT)\left[\frac{\cos(kZ)-1}{(kZ)^2} + 2\frac{kZ
-\sin(kZ)}{(kZ)^3} \right] \right\}
\end{eqnarray}
This leads to
\begin{eqnarray}
g_{00}= 1 + k^2 A_{ab}X^aX^b\left[ \cos(kT)\frac{kZ-\sin(kZ)}{(kZ)^2}
+ \sin(kT)\frac{\cos(kZ)-1}{(kZ)^2} \right] ,\\
g_{03}=k^2 A_{ab}X^aX^b\left\{ \cos(kT)\left[\frac{\sin(kZ) -
kZ}{(kZ)^2} + \frac{\cos(kZ)-1+(kZ)^2/2}{(kZ)^3}
\right]  \right.  \nonumber \\
\left.+ \sin(kT)\left[\frac{\sin(kZ) -
kZ}{(kZ)^2}-\frac{\cos(kZ)-1}{(kZ)^2}\right] \right\} ,\\
g_{0a}=- k^2 A_{ab}X^b Z\left\{ \cos(kT)\left[\frac{\sin(kZ) -
kZ}{(kZ)^2} + \frac{\cos(kZ)-1+(kZ)^2/2}{(kZ)^3}
\right]   \right.  \nonumber \\
\left.+ \sin(kT)\left[\frac{\sin(kZ) -
kZ}{(kZ)^2}-\frac{\cos(kZ)-1}{(kZ)^2}\right] \right\} ,\\
g_{ab}=\eta_{ab} + k^2 A_{ab}Z^2\left\{ \cos(kT)\left[\frac{kZ
-\sin(kZ)}{(kZ)^2} - 2\frac{\cos(kZ)-1+(kZ)^2/2}{(kZ)^3} \right]   \right.
\nonumber \\ \left.+ \sin(kT)\left[\frac{\cos(kZ)-1}{(kZ)^2} + 2\frac{kZ
-\sin(kZ)}{(kZ)^3} \right] \right\}  \\
g_{a3}=- k^2 A_{ab}X^b Z\left\{ \cos(kT)\left[\frac{kZ
-\sin(kZ)}{(kZ)^2} - 2\frac{\cos(kZ)-1+(kZ)^2/2}{(kZ)^3} \right]   \right.
\nonumber \\ \left.+ \sin(kT)\left[\frac{\cos(kZ)-1}{(kZ)^2} + 2\frac{kZ
-\sin(kZ)}{(kZ)^3} \right] \right\}  \\
g_{33}=-1 + k^2 A_{ab}X^aX^b\left\{ \cos(kT)\left[\frac{kZ
-\sin(kZ)}{(kZ)^2} - 2\frac{\cos(kZ)-1+(kZ)^2/2}{(kZ)^3} \right]
\right.\nonumber \\
\left.+ \sin(kT)\left[\frac{\cos(kZ)-1}{(kZ)^2} + 2\frac{kZ
-\sin(kZ)}{(kZ)^3} \right] \right\},
\end{eqnarray}
where we set
\begin{eqnarray}
A_{ab}X^b= A_{+}(\delta^1_a X - \delta^2_a Y)
+ A_{\times}(\delta^1_a Y + \delta^2_a X), \nonumber \\
A_{ab}X^a X^b= A_{+}(X^2 - Y^2)+2 A_{\times} XY \nonumber .
\end{eqnarray}
Our results agree with the metric obtained in \cite{F}.

{\it Geodesic deviation.} Let us consider the geodesic deviation of two
neighbouring geodesics in the gravitational field of a plane
weak wave. The separation vector $\eta$ satisfies a geodesic deviation
equation (\ref{dev1}), which reads
\begin{equation}
\frac{d^2
\eta^{(\alpha)}}{d\lambda^2}={R^{(\alpha)}}_{(\beta)(\gamma)(\delta)}
\zeta^{(\beta)}\zeta^{(\gamma)}\eta^{(\delta)},
\label{dev2}
\end{equation}
$\lambda$ being a canonical parameter along the geodesic and $\zeta$ a
tangent vector to it.
The solution of this equation is given by
\begin{equation}
\eta^{(\alpha)}= \eta^\alpha_{0}
+ {\int_0^\lambda d\tau\int_0^\tau d\tau '
{R^{\alpha}}_{\beta \gamma \delta}\stackrel{0~}{\zeta^{\beta}}
\stackrel{0~}{\zeta^{\gamma}} \eta^{\delta}_0}, \label{Int_2}
+ {\cal O}(R^2)
\end{equation}
where we set $\eta^\alpha_0 = \eta^\alpha(0)$ and integration being
performed along the basic geodesic line canonically parametrized by
$\lambda$

The separation magnitude is found to be
\begin{equation}
l= l_{0}
- {\int_0^\lambda d\tau\int_0^\tau d\tau '
{R_{\alpha \beta \gamma \delta}
\stackrel{0~}{\zeta^{\beta}}\stackrel{0~}{\zeta^{\gamma}}
\eta^{\delta}_0} \eta^{\alpha}_0},
\label{Int_3}
\end{equation}

Now let us consider the geodesic deviation of two neighbouring particles
$A$ and $B$ assuming that the the origin of Fermi reference frame attached
to $A'$s geodesic and for $T=0$ the particles are in the plane $Z=0$. In
this case the components of the separation vector are nothing but the
coordinates of particle B:  $\eta^\alpha = X^\alpha$,
$\stackrel{0~}{\zeta^\beta}=\delta^\beta_0$ and the canonical parameter
$\lambda$ being a proper time $\tau$ along the geodesic line of the
observer. Then (\ref{Int_3}) reads
\begin{eqnarray}
l= l_{0}
- \frac{1}{l_0} \int_0^\lambda d\tau\int_0^\tau d\tau '
R_{0ab0}X^a X^b .
\end{eqnarray}
This yields
\begin{eqnarray}
l= l_{0}
- \frac{1}{l_0} h_{ab}(T)X^a X^b
\label{Sol}
\end{eqnarray}
and for the plane monochromatic gravitational wave one obtains the well
known result \cite{MTW}
\begin{eqnarray}
l =l_{0} - \frac{l_0}{2}\left( A_{+}\cos 2\varphi
+ A_{\times}\sin 2\varphi \right) \sin(kT).
\label{Sol0}
\end{eqnarray}

\section *{ACKNOWLEDGEMENTS}
It is a pleasure to acknowledge numerous conversations with
N. V. Mitskievich. This work was supported by CONACyT grant 1626 P-E.


\section* {Appendix A}

Here we shall obtain the formula (Eq.(\ref{exp1}) in the text)
\[
\nabla_{\partial_\lambda}{e_{(\nu)}}^\mu=
-\frac{1}{u}\int_0^u\tau d\tau{R^\mu}_{\nu \lambda\delta}\stackrel{0~}
{\xi^\delta}+\mbox{\scriptsize$\cal O$}(R).
\]

Let us start with the Taylor expansion for tetrad in the neighbourhood
$V(\em p_0)$
\begin{equation}
e_{(\nu)}{}^\mu=
\delta^\mu_\nu +\frac{{d\stackrel{0}e_{(\nu)}}^\mu}{d u}u
+\frac{1}{2!}\frac{{d^2\stackrel{0}e_{(\nu)}}^\mu}{d u^2}u^2
+\frac{1}{3!}\frac{{d^3\stackrel{0}e_{(\nu)}}^\mu}{d u^3}u^3+ \cdots ,
\label{A1}
\end{equation}
$u$ being the canonical parameter along  geodesics starting in the point
${\em p}_ 0$. Using in the Riemann normal coordinates the equation of
parallel propagation
\begin{equation}
\frac{d e_{(\nu)}{}^\mu}{d u} +
\Gamma^\mu_{\sigma\lambda }e_{(\nu)}{}^\sigma\xi^\lambda=0,
\end{equation}
we rewrite (\ref{A1}) as
\begin{equation}
e_{(\nu)}{}^\mu= \delta^\mu_\nu
-\frac{1}{2!}\stackrel{0~}{\Gamma^{\mu}}_{\nu \sigma,\gamma}X^\sigma X^\gamma
-\frac{1}{3!}\stackrel{0~}{\Gamma^{\mu}}_{\nu \sigma,\gamma,\delta}X^\sigma
X^\gamma X^\delta + \cdots + {\cal O}(R^2).
\label{A3}
\end{equation}
The expansion of the connection coefficients is given by
\begin{equation}
\Gamma^\mu_{\nu\lambda}=\stackrel{0~}{\Gamma^{\mu}}_{\nu\lambda, i}X^i
+\frac{1}{2!}\stackrel{0~}{\Gamma^{\mu}}_{\nu\lambda,i,l}X^i X^l  + \cdots .
\label{A4}
\end{equation}
Applying Eq. (\ref{A3}), (\ref{A4}) we obtain the following series for the
covariant derivatives of tetrad
\begin{eqnarray}
{\nabla_{\partial_\lambda}}e_{(\nu)}{}^\mu=
\stackrel{0~}{\Gamma^{\mu}}_{\nu \lambda,\sigma }X^\sigma  - \stackrel{0~}{\Gamma^{\mu}}_{\nu(\lambda,\sigma )}X^\sigma
+\frac{1}{2!}(\stackrel{0~}{\Gamma^{\mu}}_{\nu \lambda,\sigma ,\gamma  }
-\stackrel{0~}{\Gamma^{\mu}}_{\nu (\lambda,\sigma ,\gamma  )}) X^\sigma  X^\gamma    \nonumber\\
+\frac{1}{3!}(\stackrel{0~}{\Gamma^{\mu}}_{\nu \lambda,\sigma ,\gamma  ,\delta   }
-\stackrel{0~}{\Gamma^{\mu}}_{\nu (\lambda,\sigma ,\gamma  ,\delta   )})X^\sigma  X^\gamma   X^\delta
+ {\cal O}(R^2).
\label{A5}
\end{eqnarray}
Now using the relations
\begin{eqnarray}
\frac{1}{2}\stackrel{0}{~R^\mu}{}_{\nu \sigma  \lambda}X^\sigma =
(\stackrel{0~}{\Gamma^{\mu}}_{\nu \lambda,\sigma }
-\stackrel{0~}{\Gamma^{\mu}}_{\nu(\lambda,\sigma )})X^\sigma , \nonumber \\
\frac{2}{3}\stackrel{0}{~R^\mu}{}_{\nu \sigma  \lambda,\gamma  }X^\sigma
X^\gamma  =(\stackrel{0~}{\Gamma^{\mu}}_{\nu \lambda,\sigma ,\gamma  }
-\stackrel{0~}{\Gamma^{\mu}}_{\nu (\lambda,\sigma ,\gamma  )}) X^\sigma
X^\gamma  ,  \nonumber\\
\frac{3}{4}\stackrel{0}{~R^\mu}{}_{\nu \sigma  \lambda,\gamma  \delta }
X^\sigma  X^\gamma   X^\delta   = (\stackrel{0~}{\Gamma^{\mu}}_{\nu
\lambda,\sigma ,\gamma,\delta } -\stackrel{0~}{\Gamma^{\mu}}_{\nu
(\lambda,\sigma ,\gamma,\delta )})X^\sigma  X^\gamma   X^\delta   ,\quad
\mbox {etc} \nonumber
\end{eqnarray}
one can write the expansion (\ref{A5}) as
\begin{eqnarray}
{\nabla_{\partial_\lambda}}e_{(\nu)}{}^\mu=
\frac{1}{2}\stackrel{0}{~R^\mu}{}_{\nu \sigma  \lambda} X^\sigma
+\frac{2}{3!}\stackrel{0}{~R^\mu}{}_{\nu \sigma  \lambda,\gamma  }X^\sigma  X^\gamma
+\frac{3}{4!}\stackrel{0}{~R^\mu}{}_{\nu \sigma  \lambda,\gamma  }X^\sigma  X^\gamma   X^\delta    + \cdots
\nonumber  \\
+\frac{n}{(n+1)!}\stackrel{0}{~R^\mu}{}_{\nu \gamma  _1 \lambda,\gamma  _2, \cdots ,\gamma  _n}
X^{\gamma  _1} X^{\gamma  _2}\cdots X^{\gamma  _n}+\cdots +{\cal O}(R^2).
\nonumber
\end{eqnarray}
It is convenient to present this series in the form
\begin{eqnarray}
{\nabla_{\partial_\lambda}}e_{(\nu)}{}^\mu=
\sum_{n=0}^{\infty}\frac{d^n\stackrel{0}{(R^\mu}{}_{\nu \sigma  \lambda}
{\xi^\sigma })}{d u^n}\frac{u^{n+1}}{(n+2)(n!)} +{\cal O}(R^2)
\end{eqnarray}
Straightforward calculation yields the following integral representation of
Eq.(\ref{A5}):
\begin{equation}
\nabla_{\partial_\lambda}{e_{(\nu)}}^\mu=
-\frac{1}{u}\int_0^u {R^\mu}_{\nu \lambda \sigma }{\xi^\sigma } \tau
d\tau +{\cal O}(R^2).
\end{equation}
Integrating by parts it found to be
\begin{equation}
\nabla_{\partial_\lambda}{e_{(\nu)}}^\mu=
-\int_0^u{R^\mu}_{\nu \lambda \sigma }{\xi^\sigma }d\tau
+\frac{1}{u}\int_0^ud\tau\int_0^\tau d\tau'
{R^\mu}_{\nu \lambda \sigma }{\xi^\sigma }+{\cal O}(R^2).
\end{equation}

\section * {Appendix B}

Here we shall obtain the series (Eq.(\ref{exp}) in the text)
\begin{eqnarray}
e^{(\mu)}_0=\delta^\mu_0+\Omega^\mu{}_i
X^i+\frac{1}{2}\stackrel{0~}{R^\mu}_{ij0}X^iX^j
+\frac{1}{6}\stackrel{0}{R}_{0ij0}\Omega^\mu{}_k
X^i X^jX^k+\frac{1}{6}\stackrel{0~}{R^\mu}_{ij0,k}X^i X^jX^k+\cdots,
\nonumber   \\
e^{(\mu)}_p=\delta^\mu_p+\frac{1}{6}\stackrel{0~}{R^\mu}_{ijp}X^iX^j
+\frac{1}{12}\stackrel{0~}{R^0}_{ijp}\Omega^\mu{}_k X^iX^jX^k
+\frac{1}{12}\stackrel{0~}{R^\mu}_{ijp,k} X^iX^jX^k+\cdots.
\end{eqnarray}

We start with the Taylor expansion for tetrad in a world tube
surrounding the world line $\gamma$ of Fermi observer:
\begin{equation}
e^{(\nu)}_\mu= \delta^\mu_\nu
+\frac{d e^{(\nu)}_\mu{\scriptstyle (0)}}{d u} u
+\frac{1}{2!}\frac{d^2 e^{(\nu)}_\mu{\scriptstyle (0)}}{d u^2} u^2
+\frac{1}{3!}\frac{d^3 {e^{(\nu)}_\mu}{\scriptstyle (0)}}{d u^3}u^3+ \cdots,
\label{T1}
\end{equation}
$u$ being the canonical parameter along spacelike geodesics ortogonal to
$\gamma$. Using in Fermi coordinates the equation of parallel propagation
\begin{equation}
\frac{d e^{(\nu)}_\mu}{d u} - \Gamma^\sigma_{\mu i}e^{(\nu)}_\sigma\xi^i=0,
\end{equation}
we rewrite Eq.(\ref{T1}) as
\begin{eqnarray}
e^{(\nu)}_\mu= \delta_\mu^\nu + \stackrel{0~}{\Gamma^{\nu}}_{\mu i}X^i
+\frac{1}{2!}\left(\stackrel{0~}{\Gamma^{\nu}}_{\mu i,l}
+ \stackrel{0~}{\Gamma^{\lambda}}_{\mu i}
\stackrel{0~}{\Gamma^{\nu}}_{\lambda l}\right)X^i X^l  \nonumber \\
+\frac{1}{3!}\left(\stackrel{0~}{\Gamma^{\nu}}_{\mu i,l,p}
+ \stackrel{0~}{\Gamma^{\lambda}}_{\mu i,l}
\stackrel{0~}{\Gamma^{\nu}}_{\lambda p}
+ \stackrel{0~}{\Gamma^{\lambda}}_{\mu i}
\stackrel{0~}{\Gamma^{\nu}}_{\lambda l, p}
+ \stackrel{0~}{\Gamma^{\lambda}}_{\mu i}
\stackrel{0~}{\Gamma^{\sigma}}_{\lambda l}
\stackrel{0~}{\Gamma^{\nu}}_{\sigma p}\right)X^i X^l X^p
+ \cdot.
\label{T2}
\end{eqnarray}

For computing of the connection coefficients and its derivatives one needs
the spacelike geodesic deviation equation
\[
\frac{D^2\eta^\mu}{d u^2} -
R^\mu{}_{ij\nu}\xi^i\xi^j \eta^\nu =0.
\]
It is convenient to write it as
\begin{equation}
\frac{d^2\eta^\mu}{d u^2} + \frac{d\Gamma^\mu_{\nu
i}}{d u}\xi^i\eta^\nu + 2\Gamma^\mu_{\nu i}\xi^i \frac{d \eta^\nu}{d u}
+\Gamma^\lambda_{\nu i} \Gamma^\mu_{\lambda j}\xi^i\xi^j \eta^\nu
= R^\mu{}_{ij\nu}\xi^i\xi^j \eta^\nu.
\label{B4}
\end{equation}
Using deviation vectors $\eta^\mu_{(i)} = u\delta^\mu_i$
and $\eta^\mu_0 = \delta^\mu_0$, we write (\ref{B4}) as follows:
\begin{eqnarray}
\frac{d\Gamma^\mu_{p i}}{d u}\xi^i u + 2\Gamma^\mu_{p i}\xi^i
+\Gamma^\lambda_{p i} \Gamma^\mu_{\lambda j}\xi^i\xi^j u
= R^\mu{}_{ijp}\xi^i\xi^j u,  \label{B5} \\
\frac{d\Gamma^\mu_{0 i}}{d u}\xi^i
+\Gamma^\lambda_{0 i} \Gamma^\mu_{\lambda j}\xi^i\xi^j
= R^\mu{}_{ij0}\xi^i\xi^j.
\label{B6}
\end{eqnarray}
From Eq.(18) one obtains $\stackrel{0~}{\Gamma^{\lambda}}_{\mu\nu} =
\delta^0_\mu \Omega^\lambda{}_\nu$. These conditions together with
(\ref{B5}), (\ref{B6}) yield
\begin{eqnarray}
\stackrel{0~}{\Gamma^{\lambda}}_{\mu\nu}
= \delta^0_\mu \Omega^\lambda{}_\nu, \quad
\frac{d \stackrel{0~}{\Gamma^{\mu}}_{ij}}{du}\xi^j
=\frac{1}{3} \left(\stackrel{0}{~R^\mu}{}_{jki}\right)\xi^j\xi^k ,
\label{B7} \\
\frac{d \stackrel{0~}{\Gamma^{\mu}}_{0i}}{du} \xi^i
= \left(\stackrel{0}{~R^\mu}{}_{ik0}
- \Omega^0{}_i\Omega^\mu{}_k \right)\xi^i\xi^k, \\
\frac{d^2 \stackrel{0~}{\Gamma^{\mu}}_{oi}}{du^2}\xi^i
=\left(\stackrel{0}{~R^\mu}{}_{ij0,k}
- \stackrel{0}R_{0ij0}\Omega^\mu{}_k
-\stackrel{0}{~R^\mu}{}_{ij0} \Omega_{0k} \right. \nonumber \\
\left. -\frac{1}{3}\stackrel{0}{~R^\mu}{}_{ijp} \Omega^p{}_k
+2\Omega^\mu{}_i\Omega_{0j}\Omega_{0k}
\right)\xi^i\xi^j\xi^k, \\
\frac{d^2 \stackrel{0~}{\Gamma^{\mu}}_{ij}}{du^2}\xi^j
=\left(\frac{1}{2} \stackrel{0}{~R^\mu}{}_{jki,p}
- \frac{1}{6} \stackrel{0}R_{0jki}\Omega^\mu{}_p \right)\xi^j\xi^k\xi^p.
\label{B8}
\end{eqnarray}
Finally, applying (\ref{B7} - \ref{B8})  and (\ref{T2}) we compute
\begin{eqnarray}
e^{(\mu)}_0=\delta^\mu_0+\Omega^\mu{}_i
X^i+\frac{1}{2}\stackrel{0~}{R^\mu}_{ij0}X^iX^j \nonumber \\
+\frac{1}{6}\stackrel{0}{R}_{0ij0}\Omega^\mu{}_k
X^i X^jX^k+\frac{1}{6}\stackrel{0~}{R^\mu}_{ij0,k}X^i X^jX^k+\cdots,
\nonumber   \\
e^{(\mu)}_p=\delta^\mu_p+\frac{1}{6}\stackrel{0~}{R^\mu}_{ijp}X^iX^j
+\frac{1}{12}\stackrel{0~}{R^0}_{ijp}\Omega^\mu{}_k X^iX^jX^k
\nonumber \\
+\frac{1}{12}\stackrel{0~}{R^\mu}_{ijp,k} X^iX^jX^k+\cdots.
\nonumber
\end{eqnarray}

\end{document}